\documentclass{article}
\usepackage[utf8]{inputenc}
\usepackage{upgreek}
\usepackage{mathrsfs}
\usepackage{amsthm}
\usepackage{mathtools}
\usepackage{physics}
\usepackage{graphicx}
\usepackage{caption}
\usepackage{float}
\usepackage{hyperref}
\usepackage[margin=1.15in]{geometry}
\usepackage{cite}
\usepackage{color}
\usepackage{titlesec}
\usepackage{amssymb}

\titleformat*{\section}{\large\bfseries}

\title{Monochromatic Mass Spectrum of Primordial Black Holes}
\author{Matthew Kleban and Cameron Norton \\ 
\emph{\small{
Center for Cosmology and Particle Physics, New York University, New York, NY}}}
\date{\today}

\begin{document}

\maketitle

\begin{abstract}
    During slow-roll inflation, non-perturbative transitions can produce bubbles of metastable vacuum.  These bubbles expand exponentially during inflation to super-horizon size, and later collapse into black holes when the expansion of the universe is decelerating.   
 Estimating the rate for these transitions during a time-dependent slow-roll phase requires the development of new techniques.  Our results show that in a broad class of models, the inflationary fine-tuning that gives rise to small density fluctuations   causes these bubbles to appear only during a time interval that is  short compared to the inflationary Hubble time.  As a result, despite the fact that the final mass of the black hole is exponentially sensitive to the moment bubbles form during inflation,  the resulting primordial black hole mass spectrum can be  nearly monochromatic.   If the transition occurs near the middle of inflation, the mass can fall in the ``asteroid" range $10^{17}-10^{22}$g in which all known observations are compatible with  black holes comprising 100\% of dark matter.
\end{abstract}

\section{Introduction}
One of the greatest mysteries in modern physics is the nature of dark matter.  Despite accounting for over 25\% of the energy density of our universe, its nature and origin remains uncertain.  Decades of searches for weakly-interacting massive particles have so far failed to find any conclusive signal \cite{WIMPwaning}.  Axion dark matter is another interesting possibility, as these  are well-motivated beyond the Standard Model particles \cite{Marsh} and can simultaneously account for other features of our universe \cite{Bachlechner:2019vcb}.  A  different possibility is that dark matter is composed of  primordial black holes (PBHs) that formed in the early universe \cite{Green:2020jor, Escriva:2022duf}.  PBHs could be formed from Standard Model matter and radiation without any exotic particle that survives until today, although the primordial mechanism that produced them in sufficient abundance likely requires new physics.  Current observational constraints leave a 5 order of magnitude window of ``asteroid" mass black holes in which a monochromatic  spectrum of PBHs could account for all of dark matter \cite{Green:2020jor, Carr_2021}. 

A challenge for PBH dark matter is identifying a plausible mechanism for producing them in the correct abundance and with mass distribution   consistent with  observational constraints. Various PBH production mechanisms have been proposed in literature (see \cite{Carr_2021} for a review).  These include a peak in the spectrum of primordial density fluctuations \cite{HawkingCarr_1974}, first-order phase transitions \cite{CrawfordSchramm_1982}, second-order phase transitions \cite{RubinKhlopov_2000}, crossovers \cite{Escriv_2023a, Escriv_2023b}, and collapse of cosmic strings \cite{Hawking_1989}. 

Here we present a variation of the mechanism first proposed in \cite{GarrigaVilenkin_2016} and followed up in \cite{DengVilenkinGarriga_2017} and \cite{DengVilenkin_2017}, in which the quantum nucleation and expansion of vacuum bubbles or domain walls during inflation creates regions that collapse later in the evolution of the universe, forming black holes. Because the vacuum bubbles form \emph{during} inflation, their size and abundance at the end of inflation -- and the masses and quantity of black holes that eventually form -- is exponentially sensitive to when during inflation the transition occurred.  These previous works assumed that the rate of production of these objects was approximately constant during inflation, and hence predicted a very broad, power-law spectrum of PBH masses.  With some parameter choices these could account for dark matter and evade observational constraints due to the relatively low abundance in any given mass window \cite{He:2023yvl}. 

By contrast, in our analysis the transition takes place over a  fraction of an inflationary efold. 
 Despite the exponential sensitivity of the mass on the production time, the resulting PBH mass  spectrum is very close to a delta function, and the abundance can be such that PBHs constitute all dark matter.  This is consistent with observational constraints if the peak of the mass distribution lies in the ``asteroid" mass range.

A delta function-like mass distribution was also found in \cite{Liu_2020}, though this paper focused on domain walls with time-varying tension instead of vacuum bubbles.  

Another variation was studied in \cite{Kusenko_2020}, where a qualitatively different potential led to an approximately constant tunneling rate and a broad PBH mass spectrum. Very recently, an interesting alternative mechanism to produce PBHs from single-field inflation that gives a fairly narrow mass distribution was studied in \cite{Escriva:2023uko}.

\section{Bubble nucleation during inflation}
Bubbles or membranes can be produced by non-perturbative quantum effects, typically because they represent an energetically preferred state.  During inflation, these defects will expand as if in flat space until they reach the inflationary horizon size, after which they will be caught in the pseudo-de Sitter expansion and grow exponentially (regardless of their tension or the evolving energy difference with the surrounding inflationary phase). If less than one such defect is produced per Hubble volume per Hubble time, the transition will not percolate because the space expands fast enough to dilute the number density exponentially.

\begin{figure}[ht]
    \centering
    \includegraphics[scale = 0.4]{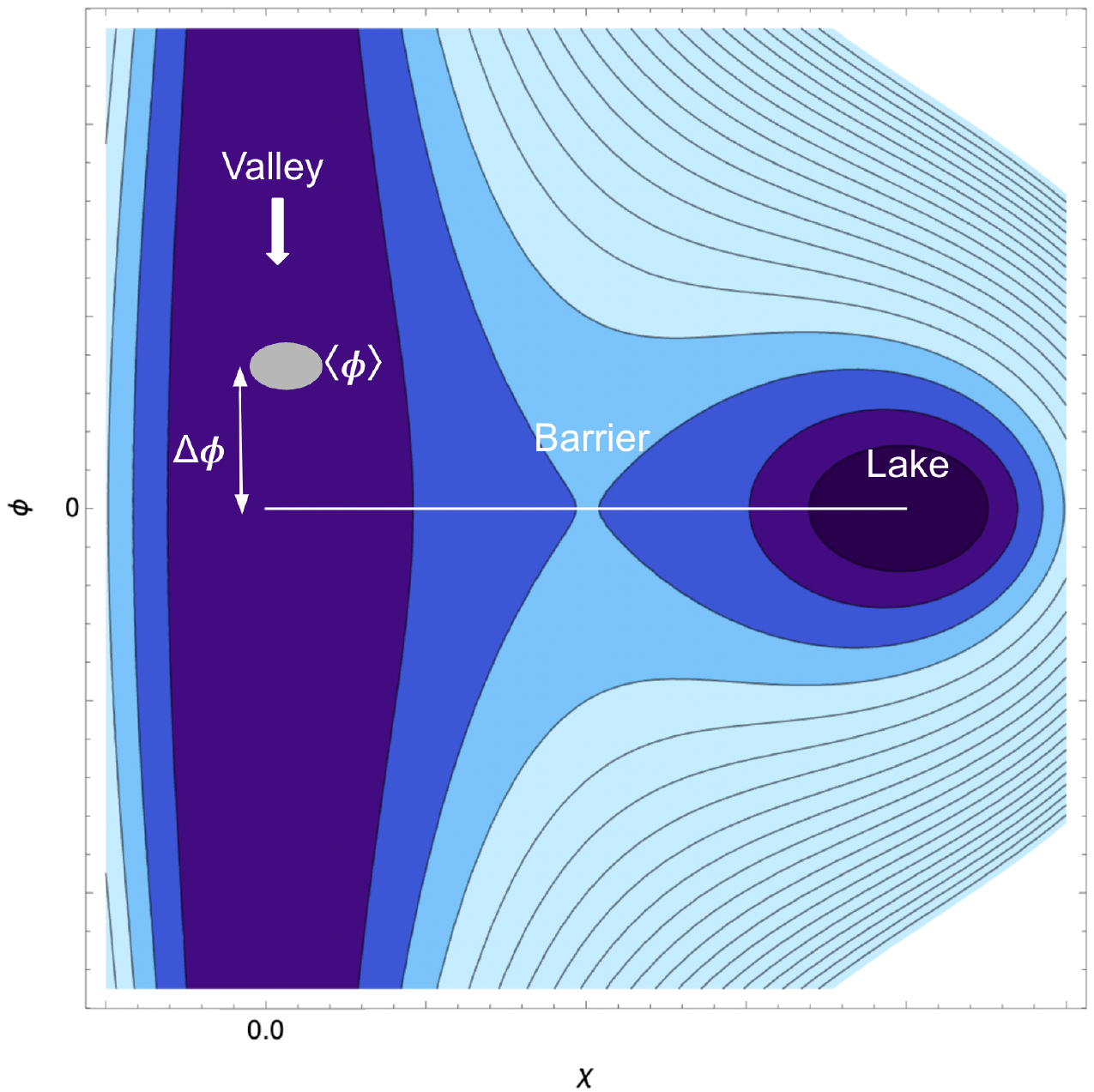}
    \caption{Schematic of the potential for a two-field inflationary model.  The inflaton $\phi$ is the vertical direction and slow-roll inflation can occur as $\phi$ evolves  downwards along a gently sloping ``valley".  The valley is separated from a  local minimum (a ``lake") by an interval in the second scalar field  $\chi$. The line indicates the (unique) instanton trajectory that connects the lake to the valley.  This instanton describes the formation of a bubble of radius $R$, inside of which the fields take values corresponding to the endpoint in the lake, and outside of which take values corresponding to the endpoint in the valley.  At a time during inflation when the vacuum expectation value $\langle \phi \rangle$ of the inflaton is displaced from the valley end-point of the instanton trajectory by a distance $\Delta \phi$, the bubble can still appear but with probability exponentially suppressed in $(\Delta \phi)^2$.}
    \label{fig:potential}
\end{figure}

For definiteness, we will assume that during inflation  two scalar fields have a potential similar to the one shown in Figure \ref{fig:potential}.\footnote{Considering a model that produces domain walls rather than vacuum bubbles would change our conclusions quantitatively, but not qualitatively.} We assume that it contains a ``valley" with a small slope (vertical direction), separated from a ``lake" by a barrier.  
Slow-roll inflation is driven by the field (labelled $\phi$) rolling  vertically down the valley in the figure. We further assume that the vacuum energy in the lake $\rho_b$ is lower than the inflationary energy density at any time during inflation, but higher than the energy density in the radiation dominated phase well after inflation ends when the vacuum bubbles re-enter the horizon.\footnote{Again, these assumptions are not necessary and could be relaxed without changing the qualitative results.}

Potentials of this form will generally admit a single unique instanton; a trajectory in field space that solves the field and gravity equations in Euclidean signature, connecting a point near the lake minimum  to a point on the other side of the barrier in the valley via a domain wall of radius $R$.\footnote{It is possible for multiple discrete instantons to exist, for instance if there are several local minima, or for a continuum of transitions to exist when there is a symmetry  or the potential for the second field is   independent of that of the first.}  Tunneling from the valley to the lake (or from the lake to the valley) corresponds to the formation of a bubble.  In the approximation that the bubble has walls thin compared to its radius, the fields inside and outside the wall will take values corresponding to the two end points of the trajectory.  One might therefore expect that as the inflaton rolls down the valley, the transition  occurs only when the inflaton wave functional assigns non-negligible probability to configurations where the inflaton field equals the valley end of the instanton trajectory in a region of  size $R$.

\subsection{Approximating the tunneling rate}

Tunneling between two local minima in the presence of gravity occurs via the Colemann-DeLuccia instanton \cite{ColemanDeLuccia_1980}.
In our case  the initial state is an inflating universe, and so we are interested in tunneling from slow-roll down the valley into the lake.  This presents an interesting complication that has not been previously studied (to our knowledge), since the initial state is time-dependent and the field is not near a minimum.  As just mentioned, for potentials of this form  there is generally a unique instanton solution that connects a specific point in the valley to the lake.  If the potential were  symmetric around $\phi = 0$ in Fig.~\ref{fig:potential}, the tunneling trajectory would lie exactly along the line $\phi=0$.  Slow roll breaks this symmetry slightly, but - absent special features or other symmetries - there is still only a single instanton trajectory.  The instanton solution for an explicit two-field potential similar to this was constructed numerically and studied in  \cite{Sasaki_2012}, where the authors were interested in tunneling from lake to valley.

We expect the tunneling rate to be maximized at the time during inflation when the vacuum expectation value (vev) of the inflaton coincides with the end point of the instanton trajectory in the valley.  Away from this time, when the field vev differs by $\Delta \phi$ from the end point of the trajectory, the rate should be  suppressed relative to this maximum.  It is important for our analysis to understand how quickly the tunneling rate goes to zero away from this maximum. To our knowledge this question has not been considered previously.  We develop two  approaches to this question that are described below.

\paragraph{Numerically estimating the rate:} One way to estimate the tunneling rate for $\Delta \phi \neq 0$ is to deform the potential slightly to create an infinitesimal potential minimum at the point in the valley from which we want to estimate the  rate. Typically this deformation creates a new instanton trajectory that connects the new minimum to the  lake.\footnote{We thank Giovanni Villadoro for suggesting this approach.} Given a specific potential, we can calculate the new instanton's action and trajectory numerically (for instance with the ``anybubble"  package \cite{Masoumi_2017}). Because the   deformation can be made arbitrarily small we expect this method to give a good approximation to the actual tunneling rate. 

\paragraph{Analytically estimating the rate:} The  numerical technique is not very informative in understanding how  the rate depends in general on $\Delta \phi$, the vertical field-space distance from the end point of the instanton. Instead, consider a less refined ``two step" analytic estimate. Starting at a given point on the valley, we  approximate the actual tunneling trajectory by a first step where the field fluctuates vertically the distance $\Delta \phi$ to the end point of the instanton trajectory, and a second step where it tunnels across the barrier via the standard Coleman-de Luccia (CdL) instanton.  The action for the full transition can be approximated as the sum of the actions for these two steps.

In order to create the initial conditions for the CdL instanton, the first fluctuation must occur in a region that is at least of size $R$, the radius of the CdL bubble.  To approximate the probability for the field to fluctuate down the valley, we calculate the variance of the field $\phi$ averaged over a sphere of radius $R$
\begin{equation}
    \phi_R (\Vec{x}, t) \equiv \frac{1}{V_3} \int_R d^3y \phi(\Vec{x} + \Vec{y}, t),
\end{equation}
where $V_3 = \frac{4}{3}\pi R^3$.
The averaged field is approximately a gaussian random variable because the inflaton is a nearly free field. The probability is therefore given by 
\begin{equation}
\exp{-\frac{(\Delta \phi)^2}{2 \sigma^2}},
\end{equation}
where $\sigma$ is the variance of the field. In our case, a simple calculation gives
\begin{equation}
    \sigma^2 = 
    \langle \phi_R (\Vec{x}, t) \phi_R(\Vec{x}, t) \rangle = \frac{9}{32\pi^2}\frac{1}{R^2},
\end{equation}
so the dependence of the probability on $\Delta \phi$ is
\begin{equation}
    \exp{-\frac{16\pi^2}{9} (\Delta \phi)^2 R^2}.
\end{equation}
Indeed, the dependence on $R^2$ and $(\Delta \phi)^2$ essentially follows from dimensional analysis.\footnote{The Euclidean action for such a fluctuation is $S \sim \int d^4 x (\partial \phi)^2 \sim  \int d^4 x (\Delta \phi/R)^2 \sim c (\Delta \phi)^2 R^2$, with associated probability $\sim e^{-S}$.} We expect the   coefficient in the exponent to be larger than $16 \pi^2/9$, because our calculation was for the \emph{average} field in a sphere of size $R$ to fluctuate $\Delta \phi$, whereas what we actually want is the more restrictive condition that the field fluctuates homogeneously everywhere in the region, in order to set up the correct initial conditions for the transition.

\begin{figure}[ht]
    \centering
    \includegraphics[scale = 0.6]{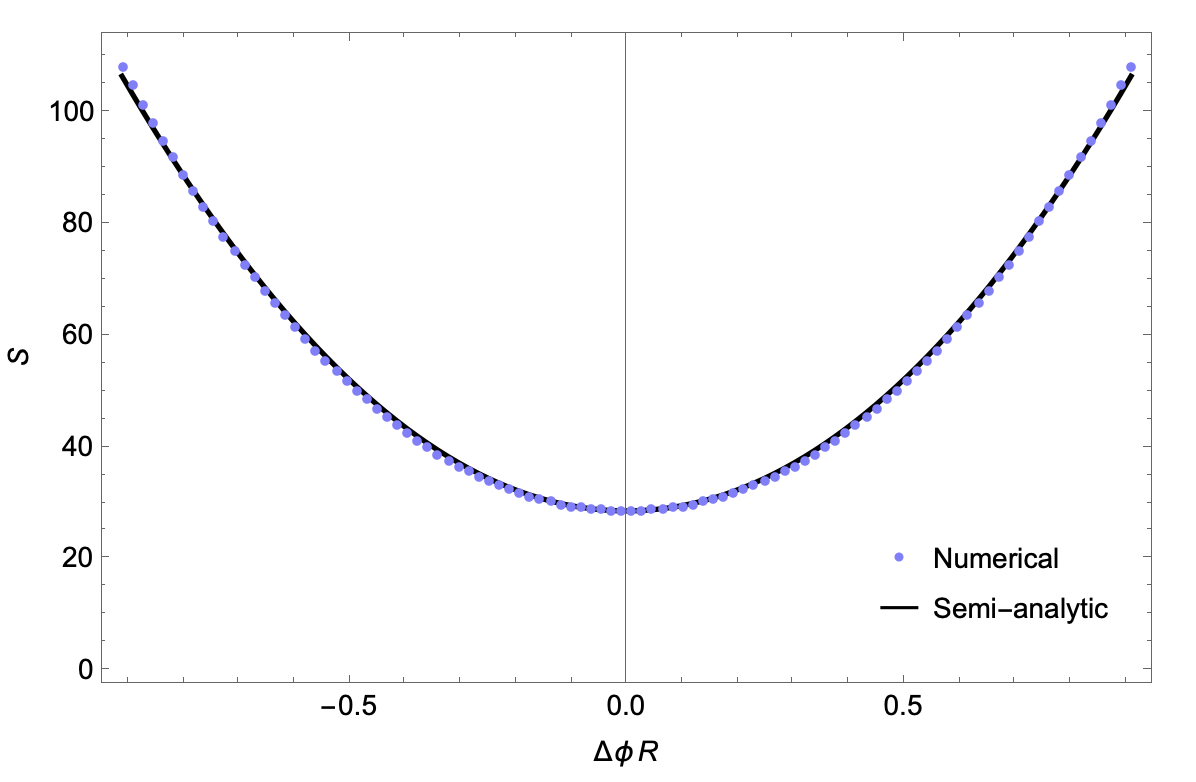}
    \caption{Action for tunneling from a point displaced  from the valley endpoint of the instanton by a distance $\Delta \phi$. \emph{Blue points:} numerical approximation $S_\text{num}(\Delta \phi)$ calculated using \cite{Masoumi_2017} from a  potential  of the form in Fig.~\ref{fig:potential}, with a small deformation added to create a local minimum when $\Delta \phi \neq 0$. (The  asymmetry in $\Delta \phi$ due to the slope of the valley is too small to be visible.)
 \emph{Black line:} Semi-analytic approximation explained in the text, $S \approx S_0 + c (\Delta \phi)^2 R^2$, where $S_0 = S_\text{num}(\Delta \phi = 0)$, $R$ is the  radius of the bubble at $\Delta \phi = 0$, and $c \approx 94 >16 \pi^2/9 $ is the best fit to the  data points shown in blue. }
    \label{Sapprox}
\end{figure}

We compare this approximation to numerical results for a specific potential using the deformation technique mentioned above  (Fig. \ref{Sapprox}).  We find that the quadratic scaling of $(\Delta \phi)^2$ in the exponent provides an excellent fit to the numerical estimates of the action made using the deformation technique, but that (as  expected) the best-fit coefficient is larger than the one in our analytic calculation.

\subsection{Slow roll}

During inflation,  an interval in the inflaton field $\Delta \phi$ is related  to an interval in the number of inflationary efolds $\Delta N$ by
\begin{equation}
    \Delta N = H_i \Delta t = \frac{H_i \Delta \phi}{\dot \phi} = \frac{H_i^2}{\dot \phi} \frac{\Delta \phi}{H_i} = 2 \pi \Delta_\mathcal{R} \frac{\Delta \phi}{H_i}.  
\end{equation}
Here $\Delta_\mathcal{R}^2$ is the power spectrum of the gauge-invariant curvature perturbation,  with $\Delta_\mathcal{R}^2 \approx 10^{-9}$ during the observable period of inflation, and we approximate the Hubble rate during inflation $H_i$ as constant. 
From our analysis in the previous subsection, we know that the transition rate is unsuppressed relative to the maximum rate when
\begin{equation}
    \Delta \phi \lesssim \frac{1}{\sqrt{c}R} \approx \frac{1}{10 R},
\end{equation}
where we have set the variance $\sigma^2 = 1/2 c R^2$ with  $c$ a dimensionless coefficient.  The last approximation uses our numerical estimate that found $c \approx 94$ (Fig.~\ref{Sapprox}).  Putting this together gives
\begin{equation}\label{DeltaN}
    \Delta N \lesssim \frac{\pi}{5} \frac{\Delta_\mathcal{R}}{H_i R}.
\end{equation}
This shows that the range of efolds during which the transition takes place is proportional to the curvature perturbation divided by the radius $R$ of the bubble measured in units of the inflationary Hubble length.  During or shortly after the observable part of inflation, the numerator $\Delta_\mathcal{R}  \approx 10^{-4.5}$ and the spectral tilt is  small and red (so that $\Delta_\mathcal{R}$ decreases slowly with time).  However, we will see that for PBHs in the asteroid mass range the transitions must take place after this phase of inflation, where we do not have a direct measurement of $\Delta_\mathcal{R}$ (and cannot be certain that the extrapolation indicated by the observed red tilt is valid).

The denominator $H_i R$ can range from $\mathcal{O}(1)$ when the bubble radius is comparable to the inflationary scale, to less than one if the bubble is  smaller.  For high-scale inflation with $H_i \approx 10^{-5} M_\text{Pl}$ we must have $H_i R \gtrsim 10^{-5}$ for the bubble to be larger than the Planck length, but for lower-scale inflationary models it is possible for $H_i R$ to be tuned smaller. 
 However, if the dynamics governing the formation of the bubble are governed by some feature of inflation it is natural  for $H_i R$ to be not much less than one.

Following the  ``step" $\Delta \phi$ that creates the initial conditions for the instanton, the field must tunnel through the potential barrier. The  tunneling rate $\lambda$ scales as 
\begin{equation}
    \lambda \sim e^{-B},
\end{equation}
where $B = S_I - S_V$ is the action of the instanton minus the action for the inflaton to stay in the valley.  Being exponentially sensitive, this rate can vary enormously.  In the next section we will calculate how large $\lambda$ should be to give the observed dark matter abundance.

\section{Vacuum bubbles and black holes}

After a vacuum bubble nucleates, pressure due to the lower energy state on the inside causes it to expand to horizon size, after which de Sitter expansion inflates it exponentially to superhorizon scales.  After inflation it continues to grow, comoving with the expansion of the universe, until eventually re-enters the horizon.  We are assuming that at this horizon-crossing time the vacuum energy inside the bubble is higher than the energy density of the radiation-dominated universe around it.  In that case the bubble begins to collapse once it re-enters the horizon.  The resulting black hole has a mass that is exponentially sensitive to the time during inflation that the bubble appeared\cite{GarrigaVilenkin_2016, DengVilenkin_2017}.

Neglecting an $\mathcal{O}(1)$ correction, the bubble's radius during inflation is approximately 
\begin{equation}
    R(t) \approx H_i^{-1}\exp[H_i(t-t_n)],
\end{equation}
where $t_n$ is the bubble nucleation time.  We denote the time of the end of inflation by $t_i$, and the radius $R(t_i) = R_i$. Letting $N_n = H_i (t_i - t_n)$ be the number of efolds before reheating that the bubble nucleates,
\begin{equation}
    R_i \approx H_i^{-1}\exp{N_n}
\end{equation}
After reheating, any initial velocity of  the bubble walls rapidly decreases due to the pressure of the fluid around the bubble, so that it expands at rest with respect to the cosmic comoving frame until it re-enters the horizon at time $t_H$ and subsequently collapses.
The mass of the resulting black hole is can be approximated as 
\begin{equation}
    G M \sim t_H, 
\end{equation}
where $t_H$ is the horizon crossing time of the co-moving scale corresponding to $R_i$.\footnote{This applies for the  case of a super-critical bubble, which we explain a bit later in this discussion.} We can find this by setting the Hubble radius equal to the radius of the bubble after inflation, 
\begin{equation}
    \frac{1}{H(t_H)} = \left(\frac{a(t_H)}{a(t_i)} \right)
\end{equation}
Assuming radiation domination, $a(t) \sim \sqrt{t}$, $t_H$ is given by
\begin{equation}
    t_H \sim \frac{R_i^2}{t_i}
\end{equation}
and the mass of the black hole as a function of $N_n$ is 
\begin{equation}
\label{mass}
    M \sim \frac{1}{G H_i} \exp{2 N_n}.
\end{equation}
(Had we considered domain walls instead \cite{GarrigaVilenkin_2016}, the mass would depend as $M \sim e^{4 N_n}$.)   Once the wall re-enters the horizon it will rapidly collapse into a black hole due to its wall tension and the fact (in the two-field model of the last section) that the vacuum inside has higher energy than the universe outside. 
A black hole of mass $M$ has a Schwarzchild radius (with  $c=1$)
\begin{equation}
    R = 2GM
 =  1.5 \times 10^{-10 }\mbox{m} \left(\frac{M}{10^{20} \mbox{g}} \right)
\end{equation}
The corresponding horizon crossing time is 
\begin{equation}
    2t_H = R 
 = 2.5 \times 10^{-19} \mbox{s} \left(\frac{M}{10^{20} \mbox{g}} \right),
\end{equation}
long before matter/radiation equality. The number of efolds before the end of inflation is
\begin{equation}
    N_n \approx 24 +  \frac{1}{2} \ln \left(\frac{M}{10^{20} \text{g}}\right)+  \frac{1}{2} \ln \left(\frac{H_i}{10^{15} \text{GeV}}\right).
\end{equation}

If the bubble expands for a time longer than its internal inflationary Hubble time before it recollapses, it will continue to inflate forever inside, forming a baby universe connected to ours through a (non-traversible) wormhole. There is a critical mass $M_{\text{cr}}$, above which a baby universe is formed and below which an ordinary black hole is formed. 
Following \cite{GarrigaVilenkin_2016}, the expression for the critical mass can be estimated as 
\begin{equation}
    G M_{\text{cr}} \sim \mbox{Min}\{t_\sigma, t_b\}, 
\end{equation}
where $t_\sigma,t_b$ are the gravitational times associated with the wall tension and vacuum energy inside the bubble.  We assume $G M_{\text{cr}} \sim t_b = H_b^{-1} \equiv \sqrt{\frac{3}{8\pi \rho_b}}$ where $\rho_b$ is the vacuum energy in the lake, so that for a bubble to be supercritical it is sufficient that
\begin{equation}
    \rho_b  > \frac{3}{8\pi G^3 M^2 } =
   \left( 3.3 \times 10^{6} \mbox{ GeV}\right)^4\left(\frac{M}{10^{20} \mbox{g}} \right)^{-2},
\end{equation}
well below the energy density in typical inflation models.  Hence, these hydrogen-atom sized PBHs contain baby universes that undergo their own internal exponential expansion and some form of decay or reheating, since the lake is at best meta-stable to further transitions.

After the black hole forms it will accrete.  It was shown in \cite{DengVilenkinGarriga_2017} that in this regime (supercritical and forming during radiation domination) the effect is to increase the mass by approximately a factor of two.  
Given the homogeneity of the early universe $\Delta_\mathcal{R} \ll 1$, we do not expect this accretion to affect the width of the black hole mass distribution significantly.

In the matter dominated phase, a some fraction of PBHs will accrete substantially due to repeated or extended encounters with stars.  We will return to this briefly in Section \ref{detect}.

We  now estimate the  spread in the mass distribution caused by the uncertainty in the nucleation time, $H \Delta t = \Delta N$ \eqref{DeltaN}. From \eqref{mass} it follows immediately that 
\begin{equation}
    \frac{\Delta M}{M} \approx 2 \Delta N. 
\end{equation}
As we have seen, it is natural for $\Delta N \ll 1$, and so the mass distribution can be very close to monochromatic.
 
\subsection{Tunneling rate}

We can now estimate the tunneling rate per Hubble volume $\lambda$ necessary to produce the observed abundance of dark matter. 
The number density of vacuum bubbles at the time they were produced is $\lambda \Delta t H_i^3$. The number density of these bubbles and the PBHs that form from them will dilute like the volume, and so at reheating the number density of bubbles is $\lambda \Delta t H_i^3 e^{-3 N_n}$, and so
the mass density of PBH dark matter today is 
\begin{equation}
     \rho_\text{PBH} \approx \lambda \Delta t M H_i^3 e^{-3 N_n}(T_0/T_\text{rh})^3.
\end{equation}
Equating this to the measured density of dark matter today and
using
 Eq. \eqref{mass} to express $N_n$ as a function of the PBH mass $M$ gives

\begin{equation}
\lambda \Delta t \approx 1.3 \times 10^{-16} \left(\frac{T_\text{rh}}{\sqrt{H_i M_\text{Pl}}}\right)^3 \left(\frac{M}{10^{20}\mbox{g}}\right)^{1/2},
\end{equation}
where $M_\text{Pl}$ is the Planck mass.  This quantity is the fraction of inflationary Hubble volumes in which a bubble nucleates during the transition.
The maximum possible reheat temperature is $T_\text{rh,max} \approx \sqrt{H_i M_\text{Pl}}$, so this is  a small number (and as a result, collisions between bubbles are very rare).  

A consistency check is that in order for the instanton approximation to be valid,  the action for the tunneling must satisfy $S \gg 1$.  The rate $\lambda = \alpha e^{-S}$.  We expect the pre-factor $\alpha$ to satisfy $\alpha \gtrsim H_i$, so we have 
\begin{equation}
S \gtrsim 25 + 3 \ln \frac{\sqrt{H_i M_\text{P}}}{T_\text{rh}} + \ln \frac{H \Delta t}{10^{-5}}  - \frac{1}{2} \ln\frac{M}{10^{20}\mbox{g}} \gg 1.
\end{equation}

\section{Constraints and detection}\label{detect}

Currently, there are no observations constraining  PBHs in the ``asteroid"  range $10^{17} \text{g} < M < 10^{23} \text{g}$ from  constituting 100\% of dark matter \cite{Green:2020jor,
Carr_2021}.  Possible approaches to detecting this form of dark matter include lensing, accumulation of one or more PBHs inside stars that affect stellar evolution over a long period of time, and stellar explosions triggered by a transit of the PBH though a star.

The lower bound on the mass range arises from Hawking radiation, which  for lighter PBHs produces gamma rays and energetic electron/positron pairs\footnote{Reference \cite{deFreitasPacheco:2023hpb} points out that if PBHs were close to extremal, the lower bound on the mass would be reduced.} \cite{Carr_2016, Boudaud_2019,Laha_2019}. These bounds could potentially be improved with future MeV telescopes or 21 cm observations \cite{Ray_2021, Saha_2022}. A study of microlensing of stars in M31   provides the upper bound on the mass range \cite{Niikura_2019}. The microscopic size of the PBHs in this range relative to optical wavelengths, combined with finite-source size effects, makes it very difficult to push theses constraints to lower PBH mass. Lensing of gamma ray bursts is of interest because  their cosmological distance and much shorter wavelength of electromagnetic radiation makes lensing by PBHs in this mass range stronger.  However, there are no current constraints from this effect \cite{Hirata_2019}.

There are two potential sources of constraints from dynamical capture of PBHs by stars: stellar survival and  observations of stellar destruction.  If a PBH passes through a star, gravitational friction heats the star and reduces the kinetic energy of the PBH.  This can lead to a bound orbit where the PBH repeatedly passes through the star, until it eventually settles into the center.  Once inside the star, the PBH will gradually accrete matter, eventually growing to the point that it strongly affects stellar evolution.

The analysis in Ref. \cite{Hirata_2019} shows that survival of stars (the observation that many stars have not been destroyed by PBHs) does not provide constraints on PBHs in the allowed mass window because captures in galaxies are rare even under optimistic assumptions. A constraint would arise only if globular clusters have high dark matter densities and low PBH velocity dispersion. Observational signatures from rare stellar destruction events present a more promising avenue for future constraints.  More modeling is needed in order to better understand the evolution and destruction of the star after the PBH is captured and accretes a substantial amount of mass (see \cite{PhysRevD.102.083004} for some recent work on neutron stars). 

Observations of white dwarfs in certain mass ranges might have implications for PBH dark matter, as PBHs could trigger an explosion via heating even in the case that they are not dynamically captured by the white dwarf, and most white dwarfs will experience at least one such transit. While an initial analysis indicated this might occur for a certain range of PBHs \cite{Graham}, a more detailed treatment shows that this process does not  provide any constraints in this mass range\cite{Hirata_2019}.

\section{Conclusion}

It is remarkable that dark matter could be composed of microscopic black holes produced in the earliest phase of the universe.\footnote{It is perhaps even more remarkable that each such atom-sized black hole contains a large universe that underwent its own period of inflationary expansion and potentially reheating and further evolution.}  The scenario considered here requires physics not far removed from what is already needed to drive inflation, without any new  forces or particle species at accessible energies.  Unfortunately, this also makes it difficult to test.

There are a number of ways in which this analysis could be extended or generalized.  One direction would be to study potentials in which the transitions occur not at one time during inflation, but at a discrete series of times.  This can be natural in inflationary models involving a pseudo-periodic potential such as unwinding inflation \cite{DAmico_2013a, DAmico_2013b, DAmico_2013c}.
We assumed that the vacuum energy in the ``lake" was well below the energy density at the end of inflation.  It would be interesting to analyze the situation where the energy density instead  falls below that  of the lake before inflation ends.  In this work we focused on the ``asteroid" mass range because of the lack of constraints on PBHs in this range.  There is another range where the constraints are weak - the so-called stupendously large BHs\cite{Carr:2020erq, Deng:2021edw}.  These black holes are larger than galactic halos and so cannot constitute all of dark matter, but evidently current constraints allow them to form an $\mathcal{O}(1)$ fraction.  The mechanism explored here can produce black holes with nearly any mass, including in this range.  It would also be of interest to extend our treatment of tunneling from slow roll to a more general analysis of tunneling from time-dependent initial states.  

\paragraph{Acknowledgements:} We would like to thank Yacine Ali-Haimoud, Heling Deng, Sergei Dubovsky, Oliver Janssen, Mehrdad Mirbabayi, and Giovanni Villadoro  for useful discussions.  Our work is supported by NSF grants PHY-1820814 and PHY-2112839.

\bibliographystyle{klebphys2}
\bibliography{references.bib}

\end{document}